\documentclass[12pt]{article}

\pdfoutput=1 
\usepackage{mathptmx}
\usepackage{graphicx}
\usepackage{siunitx}
\usepackage{amsmath}
\usepackage{amssymb}
\usepackage{amsthm}
\usepackage{times}

\title{Compressed Sensing of Field-resolved Molecular Fingerprints Beyond the Nyquist Frequency}

\author
{Kilian Scheffter$^{1,2}$, Jonathan Will$^{1,2}$, Claudius Riek$^{3}$, Herve Jousselin$^{4}$,\\ Sébastien Coudreau$^{4}$, Nicolas Forget$^{4}$, Hanieh Fattahi$^{1,2\ast}$\\
\\
\normalsize{$^{1}$Max Planck Institute for the Science of Light,}\\
\normalsize{Staudtstr. 2, 91058 Erlangen, Germany}\\
\normalsize{$^{2}$Friedrich-Alexander University Erlangen-Nürnberg,}\\
\normalsize{Staudtrstr. 7, 91085 Erlangen, Germany}\\
\normalsize{$^{3}$Zurich Instruments Germany, Mühldorfstraße 15, 81671 Munich, Germany}\\
\normalsize{$^{4}$Fastlite, rue des Cistes 165, 06600 Antibes, France}\\
\\
\normalsize{$^\ast$ E-mail:  hanieh.fattahi@mpl.mpg.de.}
}

\date{}

\begin{document} 

\maketitle 

\begin{abstract}
Ultrashort time-domain spectroscopy and field-resolved spectroscopy of molecular fingerprints are gold standards for detecting samples’ constituents and internal dynamics. However, they are hindered by the Nyquist criterion, leading to prolonged data acquisition, processing times, and sizable data volumes. In this work, we present the first experimental demonstration of compressed sensing on field-resolved molecular fingerprinting by employing random scanning. Our measurements enable pinpointing the primary absorption peaks of atmospheric water vapor in response to terahertz light transients while sampling beyond the Nyquist limit. By drastically undersampling the electric field of the molecular response at a Nyquist frequency of 0.8\,THz, we could successfully identify water absorption peaks up to 2.5 THz with a mean squared error of $12\times 10^{-4}$. To our knowledge, this is the first experimental demonstration of time-domain compressed sensing, paving the path towards real-time field-resolved fingerprinting and acceleration of advanced spectroscopic techniques.
\end{abstract}

\section{Introduction}

Detailed description of matter's constituent and internal dynamics is mirrored in its transient response to an external field. Resolving and monitoring the encoded information in the electric field of the ultrashort excitation field or in the time-dependent changes of the optical properties of the sample provides deep understanding and insights of matter \cite{wang2012molecular,coutaz2018principles}. The availability of few-cycle pulses has not only advanced various spectroscopic methods such as pump-probe spectroscopy and Fourier transform spectroscopy, but has also led to the emergence of innovative techniques such as dual-comb spectroscopy and field-resolved spectroscopy \cite{Zewail2000, Keilmann:04, Schiller:02, Picqué2019, Lanin2014, doi:10.1063/1.114909, timmers2018molecular, RevModPhys.50.607, doi:10.1080/05704920701829043, Pupeza2020, doi:10.1126/sciadv.aaw8794, Herbst_2022}.

\begin{figure}[ht]
    \centering
    \centerline{\includegraphics{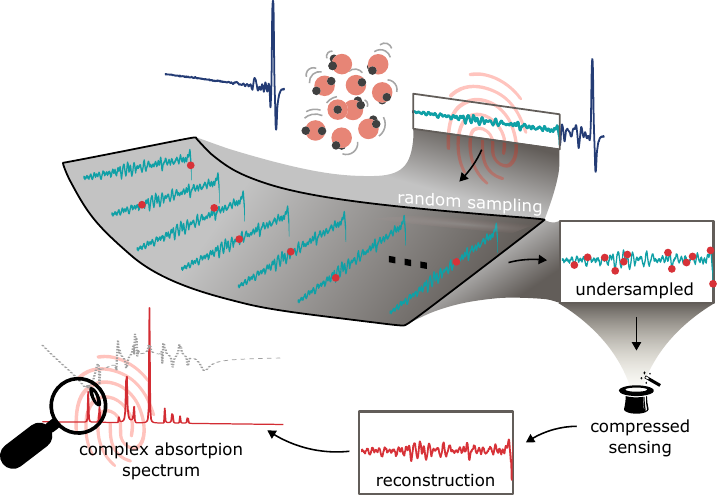}}
    \caption{Visual summary of compressed sensing of field-resolved molecular fingerprints.}
    \label{Fig.1}
\end{figure}

Employing ultrashort pulses offers two distinct advantages for spectroscopic applications (Fig. \ref{Fig.1}). Firstly, their broad spectral bandwidth enables simultaneous data acquisition of the sample, eliminating the need for repeated measurements or laser tuning. With high-bandwidth acquisition, prior knowledge of the sample is not required, as all available information can be extracted from the measurement during post-processing. Secondly, their extreme temporal confinement allows for temporal gating of the sample's response from the excitation pulses. This response, enriched with comprehensive spectroscopic information, lasts from tens of femtoseconds to nanoseconds and is commonly probed by a shorter pulse at various time intervals. When combined with additional temporal or spatial dimensions, such as multi-dimensional coherent spectroscopy \cite{smallwood}, four-dimensional imaging \cite{Cocker2016}, and hyperspectral imaging \cite{Vicentini2021}, ultrashort spectroscopic techniques render the quantitative, multivariate characterization of the sample under scrutiny and facilitate the identification of unknown constituents. However, real-time measurements are prevented due to the need to record the high bandwidth spectrum at each pixel image and time delay, which results in a prohibitively long acquisition time to attain sufficient signal-to-noise ratio. The measurement speed is limited to i) the required number of sample points dictated by the Nyquist-Shannon criteria, ii) the speed of spatio-temporal scanning, and iii) the transportation and storage speed of the measured data. Although using short, high-bandwidth excitation pulses offers significant benefits, such as simultaneous data acquisition, the sample response is often a linear combination of $K$ basis vectors leading to a $K$-sparse frequency domain. Applying algorithms to the acquisition process is necessary to minimize data redundancy and obtain the desired spectral content with the minimum required data.

According to the Nyquist-Shannon sampling theorem, for a given sampling rate, the maximum resolvable frequency of a signal, called Nyquist frequency, is defined as half the sampling rate ($f_\text{Nyquist} = f_\text{samp}/2$). This criterion imposes a boundary for the minimum required data points necessary for successfully sampling a signal. Compressed Sensing can circumvent this fundamental barrier by exploiting sparsity and incoherence of the signal, enabling the recovery of the original signal from fewer samples or measurements while preserving its quality \cite{Edgar2019,Rani2018,Candes2008,Candes2006,Donoho2006,Candes2006_2}. The prior knowledge of a signal's sparsity enables the formulation of an optimization problem that allows for the reconstruction of the signal using a reduced number of sampling points below the Nyquist-Shannon criteria. The sparsity of the measured spectroscopic information of various materials has been discussed by analytical treatment in over-sampled measurements \cite{Ostic2021,TAKIZAWA2020103042,PhysRevApplied.15.024032,Katz2010,Kaestner2018} or analytical compression before storage or transmission \cite{Kawai2021,PhysRevApplied.18.024025}. However, the requirement for incoherence and random sampling has hindered the experimental demonstration of real-time, time-domain sparse sampling. This work demonstrates real-time, field-resolved compressed sensing of water vapor molecules beyond the Nyquist criteria. The reconstruction of the absorption spectrum of water vapor sets clear boundaries on the required sample sparsity for the effectiveness of compressed sensing, as the absorption spectrum includes both high cross-section peaks and low amplitude adjacent peaks. The real-time measurement in our approach is enabled by random sampling, using a rapidly scanning delay line and a fast reconstruction algorithm for real-time data analysis.

\section{Methods}

\subsection*{Experimental setup}

\begin{figure}[h!]
    \centering
    \centerline{\includegraphics[scale=1]{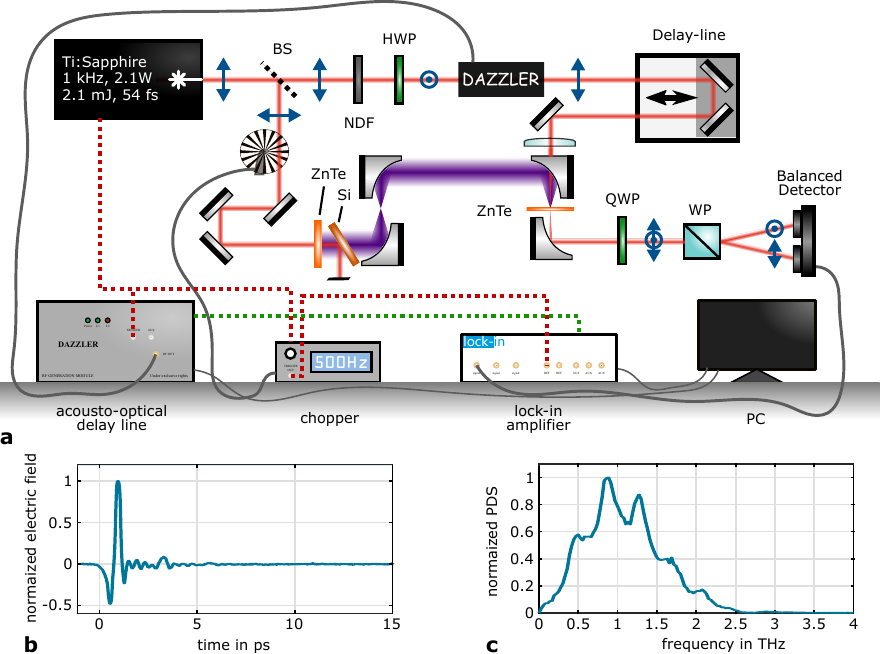}}
    \caption{a) Details of the experimental setup. b) Field-resolved measurement of the THz pulses via electro-optic sampling. c) Spectrum of the THz pulses retrieved from the time-domain measurement. BS: beam splitter, HWP: half-wave plate, NDF: neutral density filter, QWP: quarter wave plate, Si-plate: silicon-plate, ZnTe: Zinc Telluride, WP: Wollaston prism }
    \label{Fig.2}
\end{figure}

Fig. \ref{Fig.2}a shows the experimental setup for terahertz (THz) field-resolved compressed sensing. The laser delivered $\SI{54}{fs}$ pulses centered at $\SI{810}{nm}$ with $\SI{2.1}{mJ}$ energy at $\SI{1}{kHz}$ repetition rate. In the setup, the diameter of the beam was first reduced from $\SI{4.1}{mm}$ to $\SI{2.8}{mm}$ at ${1/e^2}$ via a Gallilean beam expander with the focal length of $f = \SI{-100}{mm}$ and $f = \SI{150}{mm}$. The beam was then split into two paths via a 90:10 unpolarized beamsplitter. The reflected beam containing $\SI{90}{\percent}$ of energy was chopped at a frequency of $\SI{500}{\hertz}$ via a mechanical chopper. Therefore, every second pulse was blocked by the chopper. The modulated signal, in combination with a boxcar filter, was used to eliminate systematic drifts of the measurement. The modulated beam was used to generate THz transients via optical rectification in a $\SI{1}{mm}$-thick ZnTe crystal (see Fig. \ref{Fig.2}b and c). After the THz generation, a $\SI{1.6}{mm}$ thick Silicon plate was used to filter the $\SI{800}{nm}$ pump beam while transmitting the THz-beam. The THz pulses propagated through a box filled and sealed with water vapor molecules at 50\% relative humidity. Nitrogen was used to control the humidity of the sealed box.

An electro-optic sampling (EOS) setup incorporating a 0.1 mm-thick ZnTe crystal was developed to characterize the electric field of the THz pulses. $\SI{40}{\micro\watt}$ of the amplifier's output power was used to probe the THz pulses at the EOS stage, where the polarization changes of the probe pulse due to interaction with THz electric field strength was detected in an ellipsometer incorporating balanced photo-diodes. The THz-pulses were focused on the detection crystal with a pierced parabola with the focal length of $f = \SI{50.3}{mm}$, while spatially overlapped with the probe beam.

For real-time, random scanning of the probe pulses over the water vapor's molecular response, an acousto-optical delay line with kHz scanning rates was integrated into the probe's beam path \cite{Kaplan2002, schubert2013rapid}. The acousto-optic delay line allows for arbitrary relative time delay between the THz pulse and probe pulses and shot-to-shot random scanning of the electric field. In front of the Dazzler, the polarization of the probe was flipped via a half-waveplate to provide the required input polarization. The diffracted output beam from the Dazzler has an orthogonal polarization relative to the input beam. Eventually, the probe beam was focused through a pierced parabola to the detection crystal with a plano-convex lens ($f = \SI{130}{mm}$). The THz-beam co-propagated with the probe beam with altered polarization depending on the instantaneous THz field strength via the Pockels effect in the detection crystal. After collimation of the probe beam with an off-axis parabola ($f = \SI{100}{mm}$), its polarization status was analyzed with an ellipsometric detection scheme consisting of a quarter-waveplate, a Wollaston prism, and a balanced photodetector. The signal of the balanced photodetector was fed into a lock-in amplifier using a boxcar filter to eliminate systematic drifts, while the lock-in amplifier's box-car filter ensured the signal's high bandwidth detection during the random scanning. Data acquisition was performed by the lock-in amplifier triggered by the radio frequency control signals from the acousto-optic delay line. Data acquisition trigger (green) and synchronization paths (red) are indicated in Fig. \ref{Fig.2}a by dashed lines. The time-delay module had a refresh time of $\SI{2}{ms}$ at an arbitrary temporal position within its scanning range. To capture the complete molecular response encoded in the THz field, which lasts for tens of picoseconds, a mechanical delay-line was added to the acousto-optic delay line, extending the scanning range of $\SI{6400}{fs}$ for a single acousto-optical delay line to $>\SI{40}{ps}$. Alternatively, the scanning range can be extended by coupling multiple acousto-optic modulators.

\subsection*{Reconstruction strategy}

\begin{figure}[h!]
    \centering
    \centerline{\includegraphics[scale=1]{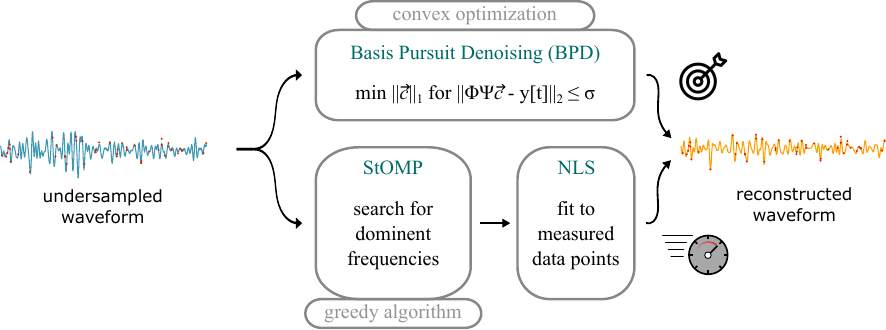}}
    \caption{The reconstruction algorithms used for the data analysis of the sparse measurements: The BPD and StOMP algorithms. While StOMP has the benefit of fast data processing, it requires an input reconstruction threshold value. To overcome this inefficiency, the IQR method was used to search for the dominant frequencies, while the NLS algorithm was used to optimize the amplitude of the reconstructed traces.}
    \label{Fig.3}
\end{figure}

Two categories of compressed sensing algorithms \cite{8260873,4472240} were investigated for reconstructing the electric field of water's randomly sampled molecular response: convex compressed sensing and greedy algorithm (see Fig. \ref{Fig.3}). Basis Pursuit Denoising (BPD) and Lasso algorithms \cite{BergFriedlander:2008,spgl1site} from the family of convex compressed sensing algorithms were used for reconstruction due to their noise robustness and low reconstruction error for signals with moderate sparsity \cite{7868430,Rani2018}. When comparing the mean squared error of the reconstruction for both algorithms, it was noted that the Lasso algorithm has a unique global minimum at a specific threshold value in the optimization problem. In contrast, the BPD algorithm demonstrates a negligible mean squared error across a range of threshold parameters ($\tau$ and $\sigma$ in supplementary information). BPD is also recognized for its ability to reduce the measurement noise of the oversampled signals \cite{doi:10.1137/S003614450037906X}. As finding a reference for arbitrary measurements to optimize the reconstruction threshold value is not always feasible, BPD was selected for further analysis. From the second category, the Stagewise Orthogonal Matching Pursuit (StOMP) greedy algorithm was chosen and developed owing to its fast computation and robustness to noise \cite{Rani2018}. To address the issue of significant amplitude reconstruction error in StOMP, a Nonlinear Least Square (NLS) algorithm has been incorporated into the StOMP reconstruction. Additionally, StOMP necessitates a distinct reconstruction threshold value at every iteration, which is inefficient and impractical. Consequently, an approach based on the Interquartile Range method \cite{dekking2005modern} was developed to calculate the threshold parameter automatically (see supplementary information for more detail).

\section{Results}
\subsection*{Real-time field-resolved compressed sensing}

As a proof of principle experiment for time domain compressed sensing, we resolved atmospheric water vapor's spectroscopic information in response to the THz excitation pulses centered at 1 THz. Among the most abundant molecules in the atmosphere, only water possesses a permanent dipole in this spectral range \cite{cox2015allen}. As a result, the spectral coverage of THz excitation pulses serves as a filter, isolating the study to only water vapor and its isotopes \cite{vanExter89}. The ambient air's absorption spectrum in this spectral range is characterized by a high density of absorption peaks \cite{CUI20153533}, which makes it an ideal platform for assessing the efficacy of compressed sensing in reconstructing moderately sparse spectra while also highlighting its limitations in reconstructing prominent absorption peaks and adjunct frequencies. Moreover, such realization holds promise for real-time gas detection in open-air environments \cite{Sitnikov_2019, 4337845}. For a comparison between conventional sampling and compressed sensing, two categories of measurements with different numbers of sampling points were performed in real-time on the water vapor response: i) linear scans with sampling points at the time intervals of $t_{n+1} = t_{n} +\Delta t$, and ii) uniformly random distributed sampling points. Each measurement was repeated ten times. As a result, the location of the randomly located sample points of the second category varies for every measurement. 

\begin{figure}[t!]
    \centering
    \centerline{\includegraphics{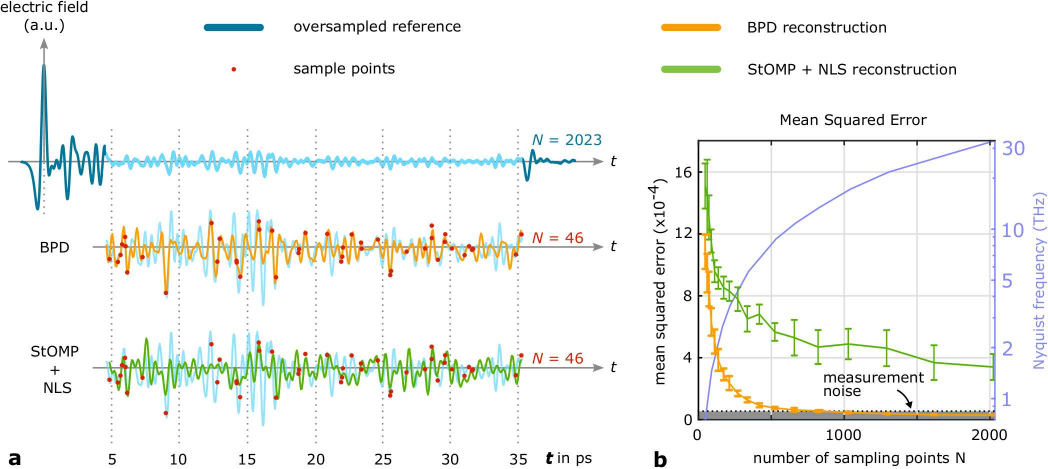}}
    \caption{a) Real-time compressed sensing of the field-resolved molecular response of atmospheric water. The light blue curve shows the reference waveform, while the red dots illustrate the random sparse detection at $N=46$ sampling points. The reconstructed waveform via the BPD and StOMP algorithms are shown in orange and green, respectively. b) Mean Squared Error of the reconstructed waveforms via BPD and StOMP versus $N$. The purple curve shows the corresponding calculated Nyquist frequency for each $N$ for comparison. The dotted line indicates the Mean Squared Error of linearly sampled measurements, indicating the measurement noise. By random sampling, the BDP filters and reduces noise from the oversampled signal. Therefore, the average of ten reconstructed waveforms shows lower noise than the case of linearly sampled measurements.}
    \label{Fig.4}
\end{figure}

Ten oversampled, linearly scanned measurements of the molecular response with $N = 2034$ sample points were averaged to generate a reference waveform. This reference trace is shown by the blue curve in Fig. \ref{Fig.4}a. The atmospheric vapor molecules are excited impulsively by the THz light transient, and their molecular response is temporarily separated from the broadband excitation pulse. This molecular response rich with spectroscopic information is sparse in the frequency domain as it only contains the information on resonance frequencies of the sample \cite{srivastava2023near}. Therefore, the temporally gated signal from $5$ to $\SI{35}{ps}$ was considered for compressed sensing. The upper limit of this range was determined by the internal reflection of the THz light transient at $\SI{36}{ps}$. To evaluate the performance of the sparse sampling, the field-resolved molecular response was measured at eighteen different sampling rates from 2023 measured data points to 46 data points. The reconstruction was performed using both BPD and StOMP algorithms.

Fig. \ref{Fig.4} a) shows the real-time compressed sensing of the field-resolved molecular response of the atmospheric water. A single sparse measurement at an extreme limit with the minimum random number of sample points of $N=46$, is shown by red dots. The orange and green curves show the reconstructed field using BPD and STOMP algorithms for $N=46$, respectively. To evaluate the performance of compressed sensing, we calculated the mean squared error between the reference waveform and the reconstructed waveforms via BPD and StOMP, respectively. 

Fig. \ref{Fig.4} b) shows the average and standard deviation of the mean squared error as a function of $N$ for each algorithm. The purple curve shows the corresponding calculated Nyquist frequency for each $N$ for comparison. The dotted line indicates the measurement noise. The measurement noise was calculated by averaging the mean squared error of the reference waveform versus the reconstructed waveform for each $N$. As this value exhibits only slight variations across different $N$, it is indicated by a horizontal dotted line. Not only does the BPD algorithm outperform the StOMP method, but for larger values of $N$, the waveforms reconstructed using BPD show a mean squared error lower than the measurement noise. This is due to the tendency of BPD to converge to a solution with maximum sparsity, which acts as a filter rejecting signal amplitudes other than absorption frequencies. For lower values of $N$, the mean squared error increases non-linearly, reaching $12\times 10^{-4}$ at $N = 46$. The low standard deviation between the ten different measurements for each value of $N$ demonstrates the stability of the reconstructions.

{Fig. \ref{Fig.5} a) displays the spectrum of the reference waveform and the sparse sampling measurements at various sampling points of $N=524$, $N=136$, and $N=46$. These sampling data points correspond to Nyquist frequencies of $\SI{8.8}{THz}$, $\SI{2.3}{THz}$, and $\SI{0.8}{THz}$, respectively. The spectra obtained from the compressed sensing reconstructed waveform successfully retrieve water absorption peaks beyond the Nyquist limit even for low sampling points of $N=46$, which corresponds to temporal steps of $\Delta t=\SI{666}{fs}$. The number of sampling points would need to be increased by at least three times to achieve a similar outcome using conventional sampling.

Fig. \ref{Fig.5} b) and c) summarize the performance of the BPD and StOMP reconstructions for measurements at various $N$, where the averaged Fourier transformation of the ten compressed sensing reconstructed waveforms at different $N$ are presented. The amplitude at each $N$ is normalized to one, and the area bordered by the purple line denotes the recovered frequencies beyond the Nyquist limit. The spectral region ranging from 1 THz to 1.5 THz is crucial for this investigation due to several absorption lines with high cross-sections and the high spectral density of the THz excitation pulses. Remarkably, the primary absorption peaks in this region are exceptionally well-recovered beyond the Nyquist limit with excellent quality. When dealing with lower values of $N$, the reconstruction of absorption lines with lower amplitude and adjunct frequencies is more noise-prone, even for those below the Nyquist limit. A comparison of panel d) and e) reveals that while the less complex StOMP algorithm has the potential of a fast reconstruction speed compared to BPD, it results in noisier reconstructions of both the low amplitude frequencies below the Nyquist limit and the absorption lines beyond the Nyquist limit.
\begin{figure}[h]
    \centering
    \centerline{\includegraphics{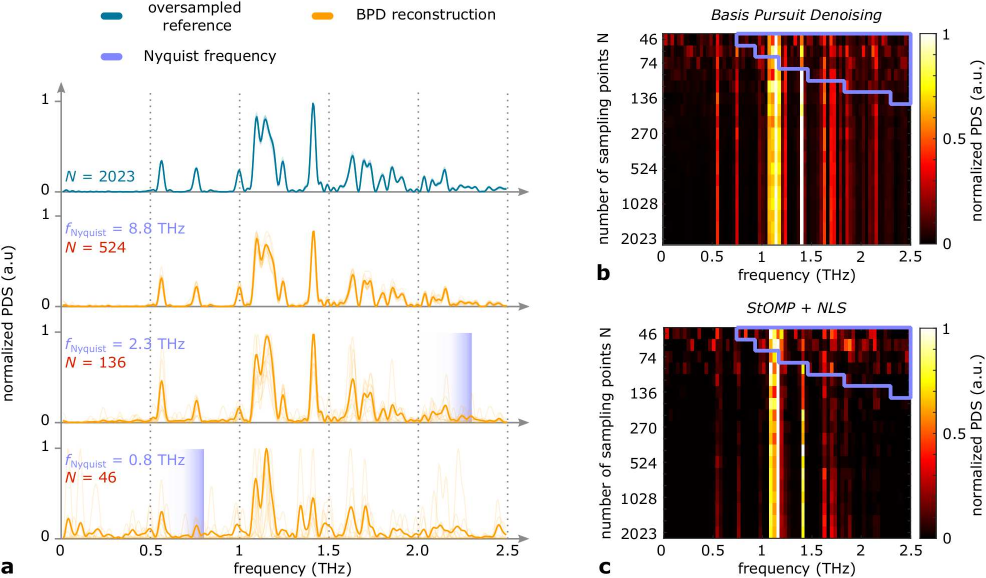}}
    \caption{a) Fourier transform counterpart of the reference waveform and the sparse sampling measurements at various sampling points of $N=524$, $N=136$, and $N=46$. The purple bars show the corresponding Nyquist frequency of each sampling data point. The transparent curve shows the variation in different random measurements. b) Averaged reconstructed spectra by BPD at various $N$. c) Averaged reconstructed spectra by StOMP at various $N$. The bordered purple area denotes the reconstructed spectral components beyond the Nyquist frequency.}
    \label{Fig.5}
\end{figure}

\section{Conclusion}

Ultrashort time-domain spectroscopy, particularly field-resolved spectroscopy, has been a gold standard for accurately identifying the constituents and dynamics of matter without the need for labels \cite{riek2015direct, RevModPhys.81.163, srivastava2023near, herbst2022recent, pandey2023ultrafast, fattahi2016sub}. Despite numerous attempts to increase measurement speed in ultrafast spectroscopy \cite{Weigel2021-kh,schubert2013rapid, Mohler:17}, real-time measurements remain challenging due to a prolonged acquisition time, significant data volume, and processing time. This study showcases compressed Sensing with rapidly sampled field-resolved spectroscopy for the first time, providing a solution to overcome these limitations. In particular, when short excitation laser pulses interact with matter, the signal-carrying information on this interaction is temporally separated from the main pulse, turning the characterization problem into a compressed sensing problem. Crucial for compressed sensing is the possibility of random sampling, which has been achieved by employing an acousto-optic delay line and a box-car filter for broadband data acquisition with a high dynamic range. The relatively high density of absorption peaks of ambient air water in the terahertz spectral range provides an ideal platform for assessing the efficacy of compressed sensing in reconstructing moderately sparse spectra while highlighting its limitations. We report resolving the absorption frequencies three times higher than the Nyquist limit. By employing a cascaded acousto-optic delay line, the acquisition time of sparse sampling for $N = 46$ sampling points can be reduced to sub-$\SI{100}{ms}$. The acquisition time is at least three times higher for the linear sampling of the signal at a Nyquist frequency of $\SI{2.3}{THz}$. Additionally, we demonstrate that compressed sensing below the Nyquist criteria can suppress measurement noise, making it valuable for speeding up measurement time and denoising sensitive measurements.

Determining when sufficient sample points have been acquired for measuring an unknown spectrum is a major challenge in compressed sensing. However, the uniform random sampling over the whole region of interest and real-time analysis of the measurement in less than $\SI{3}{ms}$ allows for the identification of the best-reconstructed waveform. As the refresh time of the acousto-optic delay line is at $\SI{30}{kHz}$, individual random shot-to-shot sampling can be performed for laser pulses below this repetition rate. For higher repetition rates, partial random scanning can be performed. Here each launched acoustic wave packet inside the acousto optical delay line delays an incoming pulse train with equally spaced delays \cite{schubert2013rapid}, while the relative delay between different scans is randomized, showing promise to introduce compressed sensing.

Real-time ultrafast spectroscopy is of crucial importance in various fields. Our innovative technique can greatly accelerate data acquisition in ultrafast spectroscopy, particularly in higher-dimensional analyses, by data volume minimization, signal acquisition time reduction, and a contraction in the required number of measurements in each dimension. These advances alleviate the requirements for specialized measurement instruments, offering benefits extending well beyond traditional spectroscopy applications. For instance, simplifying the handling of fragile specimens, enabling real-time environmental monitoring of short-lived pollutants, and real-time, open-air diagnostics of toxic and hazardous gases \cite{smallwood, Cocker2016, Vicentini2021, jun2023highly}.

\section*{Acknowledgments}
The authors thank Philip J. Russell, Francesco Tani, Martin Butryn, Stefan Malzer, and Heidi Potts for their support. 

\section*{Author Contributions}
H.F. envisioned the experiment. K.S. performed the measurements. H.J, S.C., N.F., K.S., H.F. designed the random scanning. C.R. contributed to devising the strategy for data acquisition. K.S., J.W., H.F. performed the data reconstruction and analysis. All authors reviewed and contributed to the manuscript text.

\section*{Funding}
This work was supported by research funding from the Max Planck Society. K.S. is part of the Max Planck School of Photonics, supported by the German Federal Ministry of Education and Research (BMBF), the Max Planck Society, and the Fraunhofer Society.  

\section*{Disclosures}
The authors declare no conflicts of interest. 
\bibliography{scibib}
\bibliographystyle{ieeetr}
\end{document}